\documentclass[prd,aps,eqsecnum,floatfix,nofootinbib,preprint,tightenlines]{revtex4}
\usepackage{latexsym}
\usepackage{amsmath}
\usepackage{hyperref}
\def\bibi{\bibitem}

\def\a{\alpha}

\def\d{\delta}
\def\e{\epsilon}                % Also, \varepsilon
\def\f{\phi}                    %       \varphi
\def\g{\gamma}

\def\l{\lambda}
\def\m{\mu}

\def\p{\pi}                     % Also, \varpi
\def\r{\rho}                    %       \varrho
\def\s{\sigma}                  %       \varsigma
\def\t{\tau}

\def\z{\zeta}
\def\D{\Delta}

\def\O{\Omega}

\def\S{\Sigma}

\def\cbo{{\,\raise-.15ex\Sc [\,}}                       % curly "
\def\svev#1{\left\langle #1\right\rangle}       % variable < >
\def\ddt#1{{\buildrel {\hbox{\LARGE .\kern-2pt.}} \over {#1}}}% double dot-over
\def\tstyle{\textstyle}

\def\ie{\mbox{\it i.e.}}

\def\tr{{\rm tr}\,}
\def\Tr{{\rm Tr}\,}

\def\half{{1\over 2}}

\def\det{{\rm det}}

\newcommand{\nn}{\nonumber}

\def\ie{{\it i.e.}}

\def\Sdet{{\rm sdet}}
\def\Tr{{\rm Tr}}
\def\str{{\rm str}}
\def\tr{{\rm tr}}
\def\Exp{{\rm exp}}
\def\I{{\bf 1}}
\def\O{{\bf 0}}
\def\ahat{{\hat a}}

\def\rhovperp{{\r_v^{\perp}}}

\def\qbar{{\overline{q}}}

\def\tq{{\tilde{q}}}
\def\tu{{\tilde{u}}}
\def\td{{\tilde{d}}}

\def\omegabar{{\overline{\omega}}}
\def\phihat{{\hat{\phi}}}
\def\SV{\Sigma_{\rm vac}}
\def\Sg{\Sigma_{\rm g}}
\def\Sq{\Sigma_{\rm q}}
\def\Sss{\S_{ss}}
\def\Svv{\S_{vv}}
\def\Ssv{\S_{sv}}
\def\Svs{\S_{vs}}

\def\mv{{m_{v}}}
\def\ms{{m_{s}}}
\def\mg{{m_{g}}}

\def\pizss{\pi^0_{ss}}
\def\pipss{\pi^+_{ss}}
\def\pimss{\pi^-_{ss}}
\def\pizvs{\pi^0_{vs}}
\def\pipvs{\pi^+_{vs}}
\def\pimvs{\pi^-_{vs}}
\def\pizsv{\pi^0_{sv}}
\def\pipsv{\pi^+_{sv}}
\def\pimsv{\pi^-_{sv}}
\def\pizvv{\pi^0_{vv}}
\def\pipvv{\pi^+_{vv}}
\def\pimvv{\pi^-_{vv}}
\def\etavv{\eta_{vv}}
\def\etasv{\eta_{sv}}
\def\etavs{\eta_{vs}}
\def\etass{\eta_{ss}}

\def\qbar{{\overline{q}}}
\def\ubar{{\overline{u}}}
\def\dbar{{\overline{d}}}
\def\tq{{\tilde{q}}}

\newcommand{\ba}{\begin{array}}
\newcommand{\ea}{\end{array}}

\begin{document}
\hyphenation{fer-mio-nic per-tur-ba-tive pa-ra-me-tri-za-tion
pa-ra-me-tri-zed a-nom-al-ous}

\renewcommand{\thefootnote}{$*$}

\hfill HU-EP-10/64

\hfill SFB/CPP-10-114

\begin{center}
\vspace*{10mm}
{\large\bf Flavor symmetry breaking in lattice QCD with a mixed action}
\\[12mm]
Oliver B\"ar,$^a$ Maarten Golterman$^b$ and Yigal Shamir$^c$
\\[8mm]
{\small\it
$^a$Department of Physics
\\Humboldt University,
Berlin, Germany}
\\[5mm]
{\small\it
$^b$Department of Physics and Astronomy
\\San Francisco State University,
San Francisco, CA~94132, USA}
\\[5mm]
{\small\it
$^c$Raymond and Beverly Sackler School of Physics and Astronomy\\
Tel-Aviv University, Ramat~Aviv, 69978~Israel}
\\[10mm]
{ABSTRACT}
\\[2mm]
\end{center}
We study the phase structure of mixed-action QCD with two Wilson
sea quarks and any number of chiral valence quarks (and ghosts),
starting from the chiral lagrangian.  \textit{A priori} the effective theory
allows for a rich phase structure,
including a phase with a condensate made of sea and valence quarks.
In such a phase, mass eigenstates would become admixtures of sea and valence
fields, and pure-sea correlation functions would depend on the parameters
of the valence sector, in contradiction with the actual setup of
mixed-action simulations.
Using that the spectrum of the chiral Dirac operator has a gap
for nonzero quark mass
we prove that spontaneous symmetry breaking of the flavor symmetries
can only occur within the sea sector.  This rules out a mixed condensate,
and implies restrictions on the low-energy constants of the effective theory.

\begin{quotation}

\end{quotation}

\renewcommand{\thefootnote}{\arabic{footnote}} \setcounter{footnote}{0}

\newpage
%============================
\section{\label{Intro} Introduction}
%============================
Dynamical lattice simulations with fermions that preserve chiral
symmetry \cite{Kaplan:1992bt,Shamir:1993zy,Neuberger:1997fp} are
extremely time consuming. The numerical cost typically exceeds
simulations with Wilson or staggered quarks by one or two orders of
magnitude \cite{Kennedy:2004ae}.  For this reason so-called mixed-action
simulations have been proposed, referring to a setup with either Wilson
or staggered sea quarks, and domain-wall or overlap valence
quarks; we will refer to such valence quarks collectively as
``chiral'' quarks.  Even though mixed-action
theories are not unitary, they are widely believed to have the correct
continuum limit. A key advantage is that the
valence sector preserves chiral symmetry except for
soft breaking by mass terms.  This is particularly
beneficial for the computation of weak matrix elements.

Quite a few mixed-action simulations with staggered sea quarks have
already been performed \cite{Renner:2004ck}.  All these simulations used
the configurations generated by the MILC collaboration with
Asqtad-improved staggered fermions \cite{Bazavov:2009bb}. Exploratory
simulations with twisted-mass Wilson fermions \cite{Bar:2006zj} or
clover fermions \cite{Durr:2007ef} in the sea sector have also been reported,
and more are expected in the near future.

Mixed-action theories can be studied in chiral perturbation theory
(ChPT) at nonzero lattice spacing
\cite{Bar:2002nr,Bar:2003mh,Bar:2005tu}. In particular, the dominant
source for unitarity violations can be studied analytically. The scalar
correlator, for example, is a sensitive probe for unitarity violations
\cite{Bardeen:2001jm,Prelovsek:2004jp,Golterman:2005xa}. It has been
shown that the numerical data for the scalar correlator agrees quite
well with the predictions of ChPT
\cite{Prelovsek:2005rf,Bernard:2007qf,Aubin:2008wk}, lending support to
the validity of mixed-action ChPT in describing lattice data.

In this paper we study the phase structure of mixed-action theories
with two Wilson-like sea quarks, and any number of
chiral valence quarks (and ghosts).  We consider the ``Aoki'' or
``large cutoff effects (LCE)'' regime, where quark masses are of order
$a^2$, the lattice spacing squared.
The phase diagram for theories with Wilson fermions in  the sea
or the valence sector has been studied by various authors
\cite{Creutz:1995wf,Sharpe:1998xm,Munster:2004am,Sharpe:2004ps,%
  Golterman:2005ie},
and an interesting nontrivial phase structure has been found.  In the
two-flavor case, depending on the sign of a low-energy constant (LEC) in
the chiral lagrangian \cite{Sharpe:1998xm},
there exists either a first order phase
transition with a nonvanishing minimal pion mass or an Aoki phase
\cite{Aoki:1983qi}. The latter is characterized by the spontaneous
breaking of isospin and parity symmetries, with two of the three pions
turning massless. We expect these scenarios to be present in the
mixed-action theory as well, but {\em a priori} the phase structure
might be more complicated.

To make the discussion more concrete, we begin with the effective chiral theory
for the case of two valence quarks.  We find that, indeed,
the potential of the effective theory allows for a richer phase structure.
In particular, depending on the sign of a certain linear combination of LECs
there is a ``mixed'' phase, characterized by a mixed condensate
built out of a sea and a valence quark.\footnote{
  As in the Aoki phase, this condensate breaks parity.}

A mixed phase of the effective chiral theory immediately raises a paradox:
The mixed condensate spontaneously breaks the separate sea and valence
flavor symmetries to the diagonal subgroup.  As a result, the mass
eigenstates are admixtures of sea and valence fields.
The masses themselves depend on both the sea and the valence quark masses.
This means, for example, that the two-point function
of pure-sea pion fields would become a superposition of exponentials,
all of which depend on the valence quark mass.  This putative situation
is clearly inconsistent with the very setup of mixed-action theories.
Indeed, in any numerical simulation, by construction the valence and ghost
determinants exactly cancel out, and pure-sea physics cannot possibly
depend on the presence of valence and ghost sectors.

In order to resolve this conundrum we turn to the underlying theory,
mixed-action QCD.
The chiral Dirac operator of the valence sector has a gap for nonzero
valence mass.  Following the arguments of Ref.~\cite{Vafa:1983tf},
we prove that none of the flavor
symmetries of the valence-ghost sector can be broken spontaneously,
for any number of valence quarks.
Since a mixed condensate would necessarily break the valence-ghost
flavor symmetry group, this rules out a mixed condensate.

While showing that all is well in the underlying theory, this state
of affairs calls into question the reliability of the effective
chiral theory, since the latter appears to allow for a mixed condensate.
Specializing once more to the case of two valences quarks,
we derive a mass inequality in the underlying theory
that constrains the mass of mixed charged pions to be not below
the smaller of pure-sea and pure-valence charged pion masses.\footnote{
  For a review of mass inequalities in standard QCD,
  see Ref.~\cite{Nussinov:1999sx}.}
In the effective theory,  this mass inequality implies
an inequality on a linear combination of LECs.
The latter excludes the region in the phase diagram in which a mixed phase
would occur, thereby preventing the effective theory
from making predictions that are inconsistent with the underlying theory.

In conclusion, the constrained effective theory
gives a consistent description of the possible phase diagram
of mixed-action QCD.

In Sec.~\ref{setup} we introduce the effective potential for mixed-action QCD
for the case of two chiral valence quarks, and list its symmetries.
In Sec.~\ref{PatternSSB} we study patterns of spontaneous symmetry breaking
at the level of the effective theory, focusing on the mixed condensate.
Some technical details are relegated to the appendices.
The main results are derived in Sec.~\ref{inequality}, and Sec.~\ref{Conclusion}
offers our concluding remarks.

%============================
\section{\label{setup} The effective potential for mixed-action QCD}
%============================
The chiral effective lagrangian for the
mixed-action theory with Wilson sea and chiral valence quarks has been
constructed in Refs.~\cite{Bar:2002nr,Bar:2003mh}.\footnote{
  For an introduction, see for example Ref.~\cite{Golterman:2009kw}.}
It is written in terms of the nonlinear field
\begin{equation}
\label{Sigma}
\S=\left( \begin{array}{cc}
  \Exp\left(\frac{2i}{f}\phi\right)&\omegabar\\
  \omega&\Exp\left(\frac{2}{f}\phihat\right)
\end{array} \right)\ .
\end{equation}
Specializing to the case of two sea and two valence quarks,
$\phi$ is a four-by-four
hermitian matrix, $\phihat$ a two-by-two hermitian matrix, while
$\omegabar$ and $\omega$ are four-by-two and two-by-four matrices,
respectively, with
Grassmann-valued entries.\footnote{The fact that the two-by-two matrix
  $\Exp\big(2\phihat/f\big)$ is not unitary but hermitian
  follows from a detailed study of the symmetries in the ghost sector
  \cite{Damgaard:1998xy,Golterman:2005ie,Sharpe:2001fh}.}
We will label
the rows and columns of $\S$ as $u_s$, $d_s$, $u_v$, $d_v$, $\tu_v$ and
$\td_v$, where $u$ stands for up, $d$ stands for down, and $s$, $v$
stand for sea, valence, respectively, while the tilde indicates ghost
quarks.

To the order we are working here the potential is \cite{Bar:2003mh}
 \begin{eqnarray}
\label{V1}
V&=&-\frac{f^2}{8}\; \str\left(\hat{m}\S^{-1}+\S \hat{m}\right)-\frac{f^2}{8}\;
\ahat\;\str\left( P_S\S^{-1}+\S P_S\right)\nn \\
&& -\ahat^2 W_M\;\str\left( T_3\S T_3\S^{-1}\right)
- \ahat^2 W_8'\;\str\left( P_S\S^{-1}P_S\S^{-1}
+\S P_S\S P_S\right) \nonumber\\
&& -\ahat^2 W_6'\;\left(\str\left( P_S\S^{-1} + \S P_S\right)\right)^2
-\ahat^2 W_7'\;\left(\str\left( P_S\S^{-1} - \S P_S\right)\right)^2 \ ,
\end{eqnarray}
where $\str$ denotes the supertrace.
Note that we need $\S^{-1}$ instead of $\S^\dagger$,
because the ghost part of this nonlinear field is not unitary.
Furthermore, we have that
$\Sdet(\S)=1$, with $\Sdet$ the super-determinant.
The parameters $\hat{m}$ and $\hat{a}$ are
proportional to the quark mass matrix and the lattice spacing, respectively:
\begin{eqnarray}
\label{BW}
\hat{m} & = & 2B_0 M\,,\qquad \hat{a} \,=\, 2 W_0 a\ ,
\end{eqnarray}
where $M$ is the real diagonal mass matrix.
The parameters $f$ and $B_0$ are the familiar LECs of continuum
ChPT, while $W_0,W_M, W'_i$ are additional LECs associated with a
nonvanishing lattice spacing. $P_S$ is the projector on the sea quark
sector, with $P_S = (1+T_3)/2$.  Explicitly,
\begin{equation}
\label{PST3}
M=\left( \begin{array}{ccc}
  \ms\I&\O&\O\\ \O&\mv\I&\O\\ \O&\O&\mg\I
\end{array} \right),\quad
P_S\,=\,\left( \begin{array}{ccc}
  \I&\O&\O\\ \O&\O&\O\\ \O&\O&\O
\end{array} \right),\quad
T_3\,=\,\left( \begin{array}{ccc}
  \I&\O&\O\\ \O&-\I&\O\\ \O&\O&-\I
\end{array} \right)\,,
\end{equation}
where $\O$ is the two-by-two null matrix and $\I$ is the two-by-two unit
matrix.  $\ms$ is the sea-quark mass, $\mv$ is the valence-quark mass,
and $\mg=|\mv|$ is the ghost-quark mass, which we always choose equal in
magnitude to the valence-quark mass.\footnote{Note that
  $\mg$ has to be taken positive in order for the QCD path integral in
  the ghost sector to be convergent.  Consistently, the sign of the
  ghost quark mass {\em cannot} be changed by a chiral symmetry
  transformation \cite{Golterman:2005ie}.\label{ftnegmv}}
Since the valence
determinant does not depend on the sign of $\mv$, the ghost determinant
cancels the valence determinant exactly for this choice.
We assume isospin symmetry in both the sea and the valence sector.

In the following, we will absorb the contribution of order $a$ in the
potential into the definition of the sea quark mass, which amounts to
shifting $\ms\to \ms+ a W_0/B_0$. This is not only convenient, but also
justified, since the order-$a$ shift contributes only to the additive
mass renormalization of the sea-quark mass.

In order to simplify the notation we rescale the potential
by $V\rightarrow Vf^2B_0/4$, and we write
\begin{eqnarray}
\label{V2}
V&=&-\str \left( M\S^{-1}+\S M\right)  -c_1\;\str\left( T_3\S T_3\S^{-1}\right)
- c_2 \;\str\left( P_S\S^{-1}P_S\S^{-1} + \S P_S\S P_S\right)\nonumber \\
&& -c_3\left(\str\left( P_S\S^{-1} + \S P_S\right)\right)^2
-c_4\left(\str\left( P_S\S^{-1} - \S P_S\right)\right)^2 \ ,
\end{eqnarray}
where the new coefficients $c_{i}$, $i=1,2,3,4$,
are proportional to the coefficients
$W_M$ and $W'_i$, $i=6,7,8$. Note that we also absorbed the factor $a^2$
into the $c_{i}$, so that these LECs are now of order $a^2$.

In writing Eq.~(\ref{V2}), we implicitly assume that all terms in $V$ are of
the same order of magnitude and equally important for the potential. In
other words, we assume that we are in the Aoki-regime
\cite{Sharpe:2004ps} with $\mv$ and the (shifted) mass $\ms$ both of
order $a^2$ in magnitude.  Terms not shown in Eq.~(\ref{V2}) are at least of
order $a^3$, $ma\sim a^3$ or $m^2\sim a^4$, with $m\sim \mv\sim \ms$.

Mixed-action QCD, and thus the effective lagrangian and its potential,
are invariant under
independent flavor rotations in the sea and in the valence sector.
The symmetry group $G$ has the structure \cite{Bar:2002nr}
\begin{eqnarray}
\label{SymGroup}
G&=& G_{\rm sea}\otimes G_{\rm val}\ ,\qquad
G_{\rm sea} \, =\,{\rm U(2)}_V\ ,\qquad G_{\rm val}={\rm U(2|2)}_V \ .
\end{eqnarray}
There is no symmetry connecting the sea and the valence
sectors, a consequence of the different fermion formulations used in each
sector in the underlying lattice theory.  In the limit of vanishing
lattice spacing this group enlarges to the ${\rm U(4|2)}_V$ symmetry of
the partially-quenched continuum theory \cite{Bernard:1993sv}.  For a
vanishing valence-quark mass the symmetry group is larger because of the
exact chiral symmetry in the valence sector. For a detailed description
of the full chiral symmetry group in the valence and ghost sectors, see
Refs.~\cite{Golterman:2005ie,Sharpe:2001fh}.

%==================================================
\section{\label{PatternSSB} Patterns of spontaneous symmetry breaking}
%==================================================
To begin our analysis, it is useful to explore the possible patterns of
spontaneous symmetry breaking.  We will first establish
that nothing interesting happens in the ghost sector, and then discuss
possible symmetry breaking patterns in the sea and valence sectors, in
order to provide a context for the results that will follow in the
subsequent section.

%======================================
\subsection{\label{vacuum} Ghost sector}
%======================================
The most general ground state has the form
\begin{equation}
\label{vev0}
\S_{\rm vac}=\left( \begin{array}{cc}
  \S_{\rm q}&\omegabar\\ \omega&\Sigma_{\rm g}
\end{array} \right)\ ,
\end{equation}
where $\S_{q}$ is a four-by-four matrix and $\S_{\rm g}$ is
a two-by-two matrix.
A question that immediately arises is whether the Grassmann parts $\omega$ and
$\omegabar$ can acquire any nonzero vacuum expectation values.
Intuitively, one
would think that this cannot happen, because a Grassmann-valued scale
does not exist.  In App.~\ref{AbsGrass} we will show that, indeed,
$\omega=\omegabar=0$.

The constraint $\Sdet(\SV)=1$ implies that both submatrices $\S_{\rm q}$ and
$\S_{\rm g}$ are regular
and that $\det(\S_{\rm q}) = \det(\S_{\rm g})$. To the order we are working,
the block-diagonal form of the vacuum expectation value $\S_{\rm vac}$
implies that the ghost and quark sectors decouple,
and the potential is a sum of two
terms, $V = V_{\rm q} + V_{\rm g}$.  The two-by-two matrix $\Sg$ is hermitian
and positive, and can be diagonalized by an isospin transformation in the ghost
sector.  This means that $\Sg$ can be written as
\begin{equation}
\label{evSg}
\Sg=\left( \begin{array}{cc} \r_1&0\\ 0&\r_2 \end{array} \right)\ ,
\end{equation}
with $\r_{1,2}>0$.  In terms of these two eigenvalues,
the ghost-sector part of the potential then becomes equal to
\begin{equation}
\label{ghostV}
V_{\rm g}= |\mv| (\r_1 + \r_1^{-1}+\r_2 + \r_2^{-1})+2c_1\ ,
\end{equation}
which  is minimized by  $\r_1=\r_2 = 1$, or, equivalently, $\Sg=\I$.

Thus, the effective theory predicts that isospin in
the ghost sector is unbroken.\footnote{
  This result generalizes to any number of flavors in the ghost sector.}
This is in agreement with the general
result we will establish in the underlying theory in Sec.~\ref{valisospin}.
As we will see next, the situation in the sea and valence sectors
is more subtle.

%==================================================
\subsection{\label{Scenarios} Scenarios in the sea-valence sector}
%==================================================
We begin by introducing a convenient parametrization of the most
general vacuum state in the sea-valence sector.
From the previous subsection
we have that $\det(\Sg)=1$,  hence $\det(\Sigma_{\rm q})=1$ as well;
$\S_{\rm q}$ is thus an element of SU(4).
For future use, we subdivide $\Sq$ into blocks of two-by-two matrices,
\begin{eqnarray}
\label{DefSq}
\Sq=\left( \begin{array}{cc}
  \S_{ss} & i\S_{sv}\\  i\S_{vs}&\Sigma_{vv}
\end{array} \right)\ .
\end{eqnarray}
Furthermore, if $\S_{vv}=\r_v\I$ is proportional to the unit matrix
(with $\r_v$ a complex number), $\Sq$ can be written in the form
\begin{equation}
\label{Sqform}
\Sq=e^{i\phi} \left( \begin{array}{cc}
  \r_v^*D^2 & i\rhovperp D \\ i\rhovperp D & \r_v \I
\end{array} \right)\ ,
\qquad D=\mbox{exp}(i\theta\t_3/2)\ ,\qquad \rhovperp=\sqrt{1-|\r_v|^2}\ ,
\end{equation}
with $\f=0$~mod~$\p/2$, $|\r_v|\le 1$, and in which
$\t_3$ is the third Pauli matrix.
The proof is given in App.~\ref{DeriveAnsatz}.

Symmetries which might be spontaneously broken by the ground state are
the flavor symmetry $G$, given in Eq.~(\ref{SymGroup}),
and the discrete symmetries parity ($P$) and charge conjugation ($C$),
under which the field $\S$ transforms according to
\begin{equation}
\label{DefPC}
\S \ \longrightarrow^{\hspace{-0.45cm}P} \ \hspace{0.2cm}\S^{-1}
\,,\qquad\quad
\S \ \longrightarrow^{\hspace{-0.45cm}C} \ \hspace{0.2cm}\S^{\rm T}\,.
\end{equation}
The vacuum state~(\ref{Sqform}) preserves charge conjugation.\footnote{
  For other orientations of the vacuum,
  charge conjugation has a more complicated form that involves
  a flavor rotation.}
Parity is broken by the Aoki condensate, which corresponds
to $\theta\ne 0$.  The mixed condensate, which corresponds
to $\rhovperp\ne 0$, or, equivalently, $|\r_v|<1$, breaks parity too.

We now turn to a more detailed discussion of possible phases.
The trivial vacuum is parametrized by $\Sss=\Svv=\I$ and $\Ssv=\Svs=\O$.
A first nontrivial example is provided by a vacuum
expectation value of the form
\begin{eqnarray}
\label{vevAoki}
\Sq=\left( \begin{array}{cc} e^{i\theta\tau_3}&\O\\ \O&\I \end{array}
    \right)\,,
\end{eqnarray}
with a nonvanishing isospin condensate in the sea sector (this corresponds
to $\r_v=1$ and $\phi=0$ in Eq.~(\ref{Sqform})).  This vacuum
state corresponds to the Aoki phase, with spontaneous breaking of parity
and flavor \cite{Aoki:1983qi}. The nonsinglet flavor group
SU(2)$_{\rm sea}$ breaks down to U(1).
Associated with this breaking are two massless
Goldstone bosons, the charged pions $\pi^{\pm}_{ss}$. The
possibility of the Aoki phase in the sea sector is expected, of course.
It has been shown in Ref.~\cite{Sharpe:1998xm} that the existence of the
Aoki phase is one of two possible scenarios for unquenched lattice QCD
with two flavors of Wilson fermions, which is precisely the sea sector
of the mixed-action theory we study here.

A new phase, unique to the mixed-action theory, would be a phase in which
$\Ssv$ and $\Svs$ are nonzero,
which corresponds to a condensate mixing sea and valence
quarks.  One way to explore whether such a phase might occur is
to consider the meson masses obtained when the potential is expanded
around the trivial vacuum, $\Sq=\I$.
Assuming $\ms,\mv \ge 0$ for the rest of this subsection,
and expanding the potential to quadratic order in $\phi$,
we find the following tree-level masses for mesons
made out of two sea quarks ($M_{ss}$),  two valence quarks
($M_{vv}$), and one sea and one valence quark ($M_{sv}$):
\begin{eqnarray}
\label{massesextended}
M_{ss}^2&=&B_0\left(2\ms+8c_2+16c_3\right)\,=\,B_0\left(2\ms+8c'_2\right)\ ,
\nonumber\\
M_{vv}^2&=&2B_0\mv\ ,\\
M_{sv}^2&=&B_0\left(\ms+\mv+4c_1+2c_2+8c_3\right)
\,=\,B_0\left(\ms+\mv+4c'_1+2c'_2\right)\ ,\nonumber
\end{eqnarray}
where
\begin{equation}
\label{primedcs}
c'_1=c_1+c_3 \ ,\qquad c'_2=c_2+2c_3\ .
\end{equation}
We note that $c_4$ does not contribute; it only contributes to the sea $\eta$
mass through a term quadratic in $\eta_{ss}$.

First, the valence meson mass vanishes when $\mv=0$, consistent with the
fact that the valence sector has exact chiral symmetry.  Second,
from the expression for $M_{ss}$, we see that spontaneous symmetry
breaking should take place in the sea sector when $c_2'<0$, and $\ms<4|c_2'|$,
because this would drive $M_{ss}^2$ to a negative value.
This is in agreement with the ChPT
argument of Ref.~\cite{Sharpe:1998xm} for the existence of an Aoki phase.

A new type of phase is suggested by the third equation of
Eq.~(\ref{massesextended}).
For $2c_1'+c_2'<0$, $M_{sv}^2$ becomes negative when $\ms+\mv$
is small enough.  The negative curvature at the origin of field space
indicates that a mixed condensate develops, alongside with mixed
Goldstone bosons.

Let us explore this possibility,
assuming that $c'_2>0$, but $2c'_1+c'_2<0$.
Since $c'_2>0$ there is no Aoki condensate in the sea sector.
Furthermore, let us assume that the vacuum takes the form
\begin{equation}
\label{Sqpaer}
\S_q=\left( \begin{array}{cc}
  \r & i\sqrt{1-\r^2}\\ i\sqrt{1-\r^2} & \r
\end{array} \right)\ ,
\end{equation}
with real $\r$. (This corresponds to choosing
all phases equal to zero in Eq.~(\ref{Sqform}).)  The effective potential
then reduces to
\begin{equation}
\label{Vsimple}
V'\equiv \frac{1}{4}V-c_1=-(\mv+\ms)\r-(2c'_1+c'_2)\r^2\ .
\end{equation}
If indeed $2c'_1+c'_2<0$, we find a minimum at
\begin{equation}
\label{mixedmin}
\r=\frac{\mv+\ms}{2|2c'_1+c'_2|}\ ,
\end{equation}
provided that the right-hand side is smaller than one
(otherwise, the minimum is at $\r=1$).

Before we turn to the mass spectrum of this mixed phase,
let us recall the familiar situation on the trivial vacuum.
The nonlinear field is parametrized as $\S=\exp((2i/f)\phi)$,
where the pseudoscalar field $\f$ is expanded as
\begin{eqnarray}
\label{newphi}
\sqrt{2}{\phi} & =&
\left(
\begin{array}{cccc}
\etass + \pizss & \sqrt{2}\, \pipss & \etavs + \pizvs &  \sqrt{2}\, \pipvs  \\
 \sqrt{2}\, \pimss & \etass - \pizss&  \sqrt{2}\, \pimvs & \etavs - \pizvs \\
\etasv + \pizsv &  \sqrt{2}\, \pipsv & \etavv + \pizvv & \sqrt{2}\,  \pipvv\\
  \sqrt{2}\, \pimsv & \etasv - \pizsv & \sqrt{2}\,  \pimvv & \etavv - \pizvv
\end{array}
\right)\,,
\end{eqnarray}
in self-explanatory notation.\footnote{For example,
$\pipvs$ is made of a valence anti-down quark and a sea up
quark, and $\pimsv$ is made of a sea
anti-up quark and a valence down quark.
Not all fields are independent due to the constraint
  $\Sdet(\S)=1$, which excludes the ``super-singlet'' field
  \cite{Bernard:1993sv} from our effective
  theory.}
Disregarding the ghost sector, the flavor symmetry consists
of the direct product
SU(2)$_{\rm sea}\, \otimes$~SU(2)$_{\rm val}\,
\otimes$~U(1)$_{{\rm sea}-{\rm val}}$.\footnote{The notation
  U(1)$_{{\rm sea}-{\rm val}}$ refers to a U(1) transformation
  with opposite phases in the sea and valence sectors.  The diagonal
  U(1)$_{{\rm sea}={\rm val}}$, where these phases are equal,
  acts trivially on meson fields.}
Pure-sea fields reside in the
upper-left two-by-two block. They transform only under SU(2)$_{\rm sea}$
and not under SU(2)$_{\rm val}$.

The mixed phase is realized by an expectation value
for neutral, mixed meson fields,
\begin{equation}
  \svev{\etavs} = \svev{\etasv} = (f/\sqrt{2})\z \ ,
  \qquad \cos\z=\r \ .
\label{mixedvev}
\end{equation}
This breaks the flavor symmetry spontaneously to the diagonal subgroup
SU(2)$_{\rm sea=val}$.
Associated with the four broken generators are the mixed Goldstone pions
\begin{eqnarray}
\label{MlessPMP}
\begin{array}{cccc}
\etavs-\etasv , & \pizvs-\pizsv ,&
\pipvs-\pipsv , & \pimsv-\pimvs .
\end{array}\
\end{eqnarray}
The nonzero-mass eigenstates are admixtures
of fields with different transformation properties under the separate
sea and valence flavor groups.
The nonvanishing masses depend on both $\ms$ and $\mv$.
The dependence is explicit, as well as implicit via $\r$ in Eq.~(\ref{mixedmin}).

Now we are facing a paradox:  If, for example, we calculate the
two-point function $\svev{\pipss(0)\pimss(x)}$,
we find that it is a superposition of exponentials coming
from the various nonzero mass eigenstates of the mixed phase.
These masses all depend on $\mv$,\footnote{
  We checked this by explicit calculation.}
as does the correlation function itself.
But this cannot possibly be correct, because, by the very setup
of mixed-action theories,
the sea sector does not depend on the valence part of the action at all.

In fact, this observation is a little more subtle than it appears, because
spontaneous symmetry breaking
takes place in the thermodynamical limit, whereas
numerical simulations are always done in finite volume.

In ChPT terminology, the analysis of the potential we have just carried out
corresponds to being in the $p$-regime for the Goldstone pions
of Eq.~(\ref{MlessPMP}).  In order to stay in the $p$-regime in finite volume,
one would have to turn on ``seeds'' for the given symmetry-breaking
pattern.  These would take the form of mixed mass terms that couple the sea
and valence quarks.  Were such mass terms to be introduced at the quark
level in the underlying theory, the valence and ghost determinants
would no longer cancel each other.  The separation into sea and valence sectors
would no longer apply, and there would
be nothing \textit{a-priori} wrong with finding that properties of
what used to be the sea sector now depend on parameters of what
used to be the valence sector.

Numerical simulations always maintain the exact cancellation
of valence and ghost determinants, because rather than having the determinants
of two types of quark cancel, the valence and ghost determinants are never
introduced in the first place.   Therefore, at this point the question arises
whether we truly have a paradox.
The answer is that the conflict between the effective and underlying
theories is a real one.  Given that no mixed mass terms ever exist
in the actual mixed-action setup, we are always
in the $\e$-regime for the Goldstone pions of Eq.~(\ref{MlessPMP}).
The correct prescription in this regime is to
first calculate Feynman diagrams using chiral perturbation theory for a
given orientation of the condensate, and then to integrate the result over
all possible orientations.  Let us return to our example of
the two-point function $\svev{\pipss(0)\pimss(x)}$, but now in finite volume.
On the vacuum~(\ref{Sqpaer}),
its leading-order (LO) value will be a superposition
of exponentials, as discussed above.
All other orientations of the vacuum may be obtained by
the combination of SU(2)$_{\rm val}$ and U(1)$_{{\rm val}}$
rotations.  These rotations
leave the operators $\pi_{ss}^\pm$ invariant, hence their two-point
function is unchanged when we integrate over all orientations
of the mixed condensate.  The finite-volume two-point function
of pure-sea pion fields would therefore depend on $\mv$
also when we are in the $\e$-regime for the Goldstone pions of Eq.~(\ref{MlessPMP}).
This prediction of the chiral effective theory is indeed in direct conflict
with the very setup of a mixed-action numerical simulation.

%==================================================
\section{\label{inequality} Possible phases}
%==================================================
If the ChPT description of QCD
with a mixed action is not to break down, there has to be some mechanism
that excludes the mixed phase of the effective theory.
This section shows that this is indeed the case:
Spontaneous symmetry breaking is entirely confined to the sea sector.

In Sec.~\ref{valisospin} we consider the valence and ghost sectors
of mixed-action QCD for arbitrary number of flavors.
Following Vafa and Witten \cite{Vafa:1983tf}, we employ a bound
on the spectrum of the chiral Dirac operator
to prove that none of the flavor symmetries
of the valence-ghost sector can break spontaneously for $\mv\ne 0$.
This rules out, in particular, a mixed condensate.

In Sec.~\ref{mixed} we study how this information is communicated
to the effective theory.  We begin by deriving a mass inequality
in the underlying theory, which constrains the mass of mixed charged pions
to be not smaller than the minimum
of the masses of pure-sea and pure-valence
charged pions.  Specializing (for technical reasons) to the case of
two flavors of valence quarks, we infer from the mass inequality another
inequality that must be satisfied by the LECs of mixed-action ChPT.
The LEC inequality, in turn, excludes the range of values
that produced the paradox of the previous section.

Finally, in Sec.~\ref{MiniPot} we use the results of the first two
subsections to conclude that the only nontrivial phase structure
occurs in the sea sector, where our analysis reduces to that of
Ref.~\cite{Sharpe:1998xm}.

%=========================================
\subsection{\label{valisospin}
Absence of flavor symmetry breaking in the valence-ghost sector}
%=========================================
In this subsection we prove
that the full valence-ghost flavor symmetry group is not broken
spontaneously for $\mv\ne 0$.
The analysis is carried out in the underlying theory, mixed-action QCD.
We first consider the valence sector alone,
and then extend the result to include the ghost sector too.

In order to avoid cumbersome notation we consider a mixed-action theory
with two chiral valence quarks $u$ and $d$,
with masses $m_u$ and $m_d$.\footnote{The proof generalizes trivially
  to any number of chiral valence quarks.}
At this point we assume
that $m_u$ and $m_d$ are both nonzero, but not necessarily equal.
The valence-sector action is
\begin{equation}
\label{ovact}
S=\ubar Du+m_u\ubar(1-{\tstyle\half}D)u+\dbar Dd
+m_d\dbar(1-{\tstyle\half}D)d\ ,
\end{equation}
where $D$ is a lattice Dirac operator satisfying the Ginsparg--Wilson relation
\cite{Ginsparg:1981bj}
\begin{equation}
\label{GW}
\{\g_5,D\}=D\g_5 D\ .
\end{equation}
Following Ref.~\cite{Vafa:1983tf}, we consider the isospin-breaking condensate
\begin{eqnarray}
\label{icond}
\left\langle\ubar(1-{\tstyle\half}D)u-\dbar(1-{\tstyle\half}D)
d\right\rangle_A&&\\
&&\hspace{-3cm}=-\frac{1}{V}\;\Tr\left((1-{\tstyle\half}D)
(D(1-{\tstyle\half}m_u)+m_u)^{-1}-[m_u\to m_d]\right) \ .
\nn
\end{eqnarray}
The subscript $A$ indicates a fixed gauge-field background.
Assuming that $D$ is $\g_5$ hermitian,
$D^\dagger=\g_5 D\g_5$, it follows from Eq.~(\ref{GW}) that $D$ is normal.
Thus, $D$ and $D^\dagger$ have a simultaneous set of eigenfunctions with
eigenvalues $\l$ and $\l^*$ respectively.
Once again using Eq.~(\ref{GW}) it follows
that $\l+\l^*=\l^*\l$, so we may write
\begin{equation}
\label{ev}
\l=1-e^{i\f}\ .%\qquad -\p<\f\le\p\ .
\end{equation}
Moreover, if $\psi$ is an eigenfunction with
eigenvalue $\l$, $\g_5\psi$ is an eigenfunction with eigenvalue $\l^*$;
hence, the eigenvalues~(\ref{ev}) come in pairs $\pm\f$ (except possibly
at the isolated points $\f=0$ or $\f=\p$).

Integrating over the gauge field,\footnote{The integration
  measure is nonnegative, see below.}
it is now straightforward to show that the
isospin-breaking condensate is equal to
\begin{eqnarray}
\label{iicond}
\left\langle\ubar(1-{\tstyle\half}D)u-\dbar(1-{\tstyle\half}D)
d\right\rangle &&\\
&&\hspace{-3cm}=-\int_{-\p}^\p d\f\;\r(\f)
\left(\frac{m_u}{m_u^2+4\tan^2{{\tstyle\half}\f}}
-\frac{m_d}{m_d^2+4\tan^2{{\tstyle\half}\f}}\right)\ ,\nn
\end{eqnarray}
where $\r(\f)$ is the spectral density.\footnote{Note that $\r(\f)=\r(-\f)$.
  For a similar expression for the chiral
  condensate with overlap fermions, see Ref.~\cite{Chandrasekharan:1998wg}.}
This clearly vanishes for $m_u-m_d\to 0$, as long as the common value
$m_u=m_d$ is nonzero.  The conclusion is that there is no
spontaneous symmetry breaking of isospin symmetry within the valence sector.

The previous analysis easily extends to the full valence-ghost sector.
Since Grassmann-valued condensates cannot occur (App.~\ref{AbsGrass}),
this leaves us to consider
a graded-symmetry breaking condensate of the form
\begin{equation}
\label{mixedvg}
\left\langle\qbar_v(1-{\tstyle\half}D)q_v+\tq^\dagger(1-{\tstyle\half}D)\tq
\right\rangle\ ,
\end{equation}
where $q_v$ is a valence quark and $\tq$ is a ghost quark.
Note the plus sign between the valence and ghost quark bilinears.
This sign is consistent with the graded symmetries in $\mbox{U}(n|n)_V$
(with $n$ the number of valence quarks),
which would be broken if this condensate
developed a nonvanishing expectation value.\footnote{
  Recall that ghost-quark fields commute with each other.
  For reference, the singlet condensate that does not break $\mbox{U}(n|n)_V$
  has a minus sign between the valence and ghost terms.}

The condensate~(\ref{mixedvg}) has a spectral representation
similar to Eq.~(\ref{iicond}), with $m_u$ on the right-hand side replaced by
the valence-quark mass $m_v$, and $m_d$ replaced by the ghost-quark mass $m_g$,
which we temporarily take to be different from
$\mv$ in order to study symmetry breaking.\footnote{
  Recall that $m_g$ is necessarily positive.
  Assuming an even number of valence quarks we may take $\mv>0$,
  because, if $\mv<0$, we have $\mv \to |\mv|$ under a nonanomalous
  chiral rotation of the valence quarks.}
Again, in the limit $m_g \to \mv$ this condensate vanishes, for the
same reasons as before, and we conclude that the full valence-ghost
flavor symmetry group is not spontaneously broken.

A corollary is that no mixed condensate can ever occur.
A bilinear sea-valence operator transforms in the fundamental representation
of the valence-ghost symmetry group.  Since this group does not
break spontaneously, bilinear sea-valence operators cannot acquire nonzero
expectation values.\footnote{
  The same argument excludes a valence-ghost Grassmann condensate.
  However, the latter was already ruled out by the general
  proof of App.~\ref{AbsGrass}.}

We end this subsection with a technical comment.
In order to probe isospin breaking in the sea sector (where it can occur),
we may have to turn on an (infinitesimal) difference $\D m$ between
the masses of the up and down sea quarks.  For nonzero $\D m$,
the two-flavor Wilson determinant is no longer positive.
However, since this only happens
for nonzero $\D m$ and at a nonzero lattice spacing,
the effect is  of order $a\D m$.  In this paper we
work to order $m \sim a^2$ only,
and so we may neglect such effects.\footnote{For a
  discussion of why the argument of Ref.~\cite{Vafa:1983tf} does not apply to
  an isospin-breaking condensate in the {\em sea} sector, in which the quarks
  fields are of the Wilson type, see Ref.~\cite{Sharpe:1998xm}.}

%==================================
\subsection{\label{mixed} Absence of mixed-phase Goldstone bosons}
%==================================
While in the previous subsection we showed that all is well
in the underlying theory, the puzzle concerning
the phase diagram of mixed-action ChPT remains to be resolved.
We will begin by deriving a mass inequality relating pure-sea, pure-valence,
and mixed pions in the underlying theory.  We will then
infer an inequality between the LECs of mixed-action ChPT,
which, as promised, excludes the region of parameters that gave rise
to the paradox of Sec.~\ref{Scenarios}.
As in previous sections, the ChPT-level analysis is restricted
to the case of two valence quarks.

We begin with the following inequality in mixed-action QCD:
\begin{equation}
\label{ineq}
\tr\left\langle\left(S^\dagger_{sd}(x,y)-S^\dagger_{vd}(x,y)\right)
\left(S_{sd}(x,y)-S_{vd}(x,y)\right)
\right\rangle\ge 0\ ,
\end{equation}
in which $S_{ik}(x,y)$, $i=s,v$ is the sea, respectively, valence quark
propagator.  The second index $k=u,d$ denotes flavor, up or down.\footnote{
  While the correlation functions under study depend on two valence flavors
  that we have conveniently denoted up and down,
  the total number of valence quarks can be any $n\ge 2$.}

Although as it stands inequality~(\ref{ineq}) depends only on the
$d$ propagator, the $u$ propagator will be encountered shortly.
In the isospin-symmetric phase, the up and down propagators are equal,
and Eq.~(\ref{ud}) below follows from $\g_5$-hermiticity of the Wilson operator.

When we get back to the effective theory, we will make use of the inequality
only in the symmetric phase.  Interestingly, only little extra effort is needed
to extend the inequality to the phase with broken isospin (the Aoki phase),
so let us make
this small detour.    In order to account for the latter possibility,
we  add a ``twisted'' mass term of the form $\m\qbar_s i\g_5\t_3 q_s$ to
the (sea) Dirac operator, where $q_s=(u_s,d_s)^T$.  This accomplishes two
things.  First, it accounts explicitly for isospin breaking.  Second, it
aligns isospin breaking along the third direction in isospin space, so
that the relevant condensate, if it forms, would be proportional to
$\qbar_s \g_5\t_3 q_s$.  We now have that
\begin{equation} \label{ud}
S^\dagger_{sd}(x,y)=\g_5 S_{su}(y,x)\g_5\ ,
\end{equation}
also when $\m\ne 0$.  Eq.~(\ref{ud}) holds in any finite volume,
and therefore also in the thermodynamical limit where $\m$ is eventually
turned off.

Using that $S^\dagger_{vd}(x,y)=\g_5S_{vu}(y,x)\g_5$
for the chiral overlap propagator as well,
inequality~(\ref{ineq}) can now be rewritten as
\begin{equation}
\label{ineq2}
G_{ss}(x,y)+G_{vv}(x,y)\ge G_{sv}(x,y)+G_{vs}(x,y)\ ,
\end{equation}
where
\begin{equation}
\label{Gdef}
G_{ij}(x,y)=
\left\langle\ubar_i(x)i\g_5d_j(x)\;\dbar_j(y)i\g_5u_i(y)\right\rangle\ .
\end{equation}
There are no disconnected contributions.  (This is true even if we are inside
the Aoki phase, because, with our choice of the twisted-mass term,
any isospin-breaking condensate must lie in the $\t_3$ direction.)

If inequality~(\ref{ineq2}) holds in mixed-action QCD, it must also hold
in the low-energy effective theory, in which $G_{ss}$ corresponds to the
sea-pion propagator, $G_{vv}$ to the valence-pion propagator, and
$G_{vs}$ and $G_{sv}$ to the mixed-pion propagators.
Technically, the translation into ChPT is done
by coupling the mixed-action QCD lagrangian to pseudoscalar sources
for the operators $\ubar_i(x)\g_5d_j(x)$ and $\dbar_j(y)\g_5u_i(y)$
\cite{Bar:2002nr,Bar:2003mh,Bar:2005tu,ABS}.
While the underlying theory is nonunitary, within the effective theory
the decay rates of the relevant correlation functions are thus interpreted
as pion masses.

The LO result can be
expressed in terms of the component fields of Eq.~(\ref{newphi}) as follows
\begin{equation}
\label{ineq3}
\langle\p^+_{ss}(x)\p^-_{ss}(y)\rangle
+\langle\p^+_{vv}(x)\p^-_{vv}(y)\rangle
\ge
\langle\p^+_{sv}(x)\p^-_{vs}(y)\rangle
+\langle\p^+_{vs}(x)\p^-_{sv}(y)\rangle\ .
\end{equation}
Here each two-point function is a tree-level (\ie, free)
propagator in the effective theory,
with mass determined by the potential, Eq.~(\ref{V1}).
For a meson of mass $M$, the tree-level  propagator is
\begin{equation}
\label{treeprop}
D(x-y) = \int\frac{d^4p}{(2\p)^4}\;\frac{e^{ip(x-y)}}{p^2+M^2}
= \frac{M}{4\p^2 r}\;K_1(Mr)\ ,
\end{equation}
in which $r=|x-y|$.  Substituting this into Eq.~(\ref{ineq3}) yields the inequality
\begin{equation}
\label{ineqfinal}
M_{ss}K_1(M_{ss}r)+M_{vv}K_1(M_{vv}r)\ge
2M_{sv}K_1(M_{sv}r)\ ,
\end{equation}
where we used that $M_{vs}=M_{sv}$.
If all masses are strictly positive, for large $r$
we can use the asymptotic behavior of $K_1$,
\begin{equation}
\label{K1large}
K_1(z)\sim\sqrt{\frac{\p}{2z}}\;e^{-z}\ ,\qquad|z|\to\infty\ ,
\end{equation}
finding
\begin{equation}
\label{massineq}
M_{sv}\ge\mbox{min}\left(M_{ss},M_{vv}\right)\ .
\end{equation}

A nontrivial consequence of the mass inequality~(\ref{massineq})
is that the LECs appearing in Eq.~(\ref{V2}), too, are subject to an inequality.
To see this, we take $\ms$ and $\mv$ large enough that
no spontaneous symmetry breaking takes place,
and the curvatures at the origin of field space,
which are given explicitly by Eq.~(\ref{massesextended}), are all positive.
We also choose $\mv$ such that $M_{vv}=M_{ss}$, which
implies that $\mv=\ms+4c'_2$.  We now find that the
mass inequality~(\ref{massineq}) translates into the LECs inequality
\begin{equation}
\label{LECineq}
2c_1-c_2=2c_1'-c_2'\ge 0\ .
\end{equation}
LECs are, by definition, independent of the quark masses.
While we have derived the inequality by considering special values
of the sea and valence masses, it must therefore hold true
for arbitrary values of $\ms$ and $\mv$.

We note that in
our example leading up to Eq.~(\ref{mixedmin}), we assumed that $c'_2>0$,
so that, with Eq.~(\ref{LECineq}), $c'_1+2c'_2>c'_1-2c'_2\ge 0$.  Therefore,
$c'_1+2c'_2<0$ never occurs, and the putative vacuum solution Eq.~(\ref{mixedmin})
is never encountered.

%==================================
\subsection{\label{MiniPot} The phase diagram}
%==================================
We now return to the ground state (Eqs.~(\ref{vev0}) and~(\ref{DefSq})),
\ie, to the minimization of the potential~(\ref{V2}), but subject to
the constraints we have inferred from the underlying theory.\footnote{
  In this paper we analyze the potential for the case of two valence quarks,
  but since the proof of Sec.~\ref{valisospin} is valid for arbitrary number
  of valence quarks, we expect that the analysis of the potential
  can be generalized accordingly.}
In Sec.~\ref{vacuum} we found that $\S_{\rm g}=\I$.
Using the result of Sec.~\ref{valisospin} we conclude that
$\S_{vv}=\I$ as well,
while using the result of both Sec.~\ref{valisospin} and Sec.~\ref{mixed}
it follows that $\Ssv=\Svs=\O$.
In addition, we know from App.~\ref{AbsGrass} that $\omega=\omegabar=0$.
Therefore, only $\Sss$ can take
on a nontrivial value, with $\det(\Sss)=1$, so that $\Sss\in\mbox{SU(2)}$.
Substituting the vacuum solution
\begin{equation}
\label{restrict}
\S_{\rm vac}=\left( \begin{array}{ccc}
  \S_{ss}&\O&\O\\ \O&\I&\O\\ \O&\O&\I
\end{array} \right)\ ,
\end{equation}
into the potential~(\ref{V2}), it reduces to
\begin{equation}
\label{reducedpot}
V=-\ms\,\tr\left(\Sss+\Sss^\dagger\right)
-\half\,c'_2\left(\tr\left(\Sss+\Sss^\dagger\right)\right)^2\ ,
\end{equation}
where $c'_2$ is defined in Eq.~(\ref{primedcs}).
This is precisely the potential  which was found in Ref.~\cite{Sharpe:1998xm}
for QCD with two Wilson flavors, and no valence sector.  Depending on the
sign of $c'_2$, either isospin and parity are broken in the sea sector
(if $c'_2<0$ and $|m_s|$ is small enough), with
$\p_{ss}^\pm$ the corresponding Goldstone bosons, or there is a first-order
phase transition (if $c'_2>0$), with a minimum nonvanishing pion mass at nonzero
lattice spacing.

In summary, adding a chiral-fermion valence sector to the dynamical
Wilson-fermion theory enlarges the flavor symmetry group,
and, in principle, allows for many new symmetry-breaking condensates.
Nevertheless, none of these condensates actually develops,
and we have recovered the usual two-flavor phase diagram for the sea sector.

%=============================
\section{\label{Conclusion} Concluding remarks}
%=============================
The chiral lagrangian for lattice QCD with a mixed action, with Wilson
sea quarks and chiral valence quarks, is rather involved.
We have studied this chiral lagrangian in the case of two sea and
two valence quarks in the LCE regime, $m \sim a^2$.
With no restrictions on the values of the order-$a^2$ LECs,
the phase structure is rather intricate.

In particular, the effective potential appears to allow for a mixed phase
with a condensate pairing valence with sea quarks.
Such a mixed phase would contradict the very setup of mixed-action
simulations, because pure-sea correlation functions would depend
on the parameters of the valence sector.

As we have shown in this paper, the underlying theory excludes
such a phase.   By extending the well-known argument of Ref.~\cite{Vafa:1983tf},
which relies on the existence of a gap in the spectrum of the chiral
Dirac operator for nonzero mass, we have shown that
none of the flavor symmetries of the valence and ghost sectors
can be broken spontaneously, for any number of valence quarks.
This forbids, in particular, a mixed condensate.

In addition, we have shown that the mixed-pion mass cannot be
smaller than the minimum of the  pure-sea and pure-valence pion masses.
This mass inequality translates into an inequality between LECs appearing
in the chiral lagrangian.  The inequality, Eq.~(\ref{LECineq}), must hold
independent of the choice of action of the underlying lattice theory.
In terms of the original LECs in Eq.~(\ref{V1}), the bound reads
\begin{equation}
2W_M - W_8'\ge 0 \ .
\label{bound}
\end{equation}
Without this restriction on the LECs,
the effective potential by itself would allow for a mixed phase.
 The inequality we found follows from expanding
the effective potential around the trivial vacuum, \ie, Eq.~(\ref{massesextended}),
and imposing the mass inequality~(\ref{massineq}).
It is possible that more constraints on these LECs exist, that would
follow from a complete study of the effective potential for arbitrary
$\S_{\rm vac}$, by imposing the constraints we found in the underlying theory.

Not surprisingly, we recover the well-known conclusion of
Ref.~\cite{Sharpe:1998xm} about the possible phase structure in the sea sector.
If $c'_2>0$, there is a first
order transition when $\ms$ changes sign, and the pion mass is always larger
than zero; if $c'_2<0$, a second order transition occurs
for small enough $\ms$,
in which isospin and parity are spontaneously broken \cite{Aoki:1983qi}.  This
raises the question whether an inequality might also be derived for $c'_2$ by
considering the charged and neutral pion masses in the sea sector.  The
reason that this is not possible, however, is that the neutral pion propagator
in QCD with broken isospin contains ``disconnected'' diagrams, so that the
arguments of Sec.~\ref{mixed} do not apply.  Note that in Sec.~\ref{mixed} we only
considered propagators for charged pions,
making sure that no disconnected contributions
appear.\footnote{Here ``disconnected'' diagrams are diagrams with disconnected
  quark loops; they are still connected if also gluon lines are taken into
  consideration.}

Finally, we expect that our conclusions generalize to other mixed
actions with a chiral valence sector.  For instance, if the sea quarks
are staggered, they might exhibit a nontrivial phase structure
\cite{Aubin:2004dm}.  In a mixed-action theory with a staggered sea
sector, this nontrivial phase structure would remain confined to the sea
sector, just as in the case we considered in this paper.

\vspace{3ex}
\noindent {\bf Acknowledgments}
\vspace{3ex}

MG thanks the Department of Physics of Humboldt Universit\"at zu Berlin,
and YS thanks the Department of Physics of San Francisco State University for
hospitality.  OB is supported in part by the Deutsche Forschungsgemeinschaft
(SFB/TR 09), MG is supported in part by the US Department of Energy,
and YS is supported by the Israel Science Foundation under grant no.~423/09.

%============
\appendix
%============
%===============================================
\section{\label{AbsGrass} Absence of Grassmann-valued condensates}
%===============================================
In this appendix we prove that Grassmann-valued condensates cannot arise.
To be concrete, we consider a possible $\svev{\qbar_s\tq_v}$,
where $q_s$ is a sea quark and $\tq_v$ a ghost quark.
The relevant part of the path integral is of the form\footnote{
  The argument for the vanishing of $\svev{\qbar_v\tq_v}$,
  with $q_v$ a valence quark, is similar, except that in this case $X=Y$.}
\begin{eqnarray}
\label{Z}
Z^{(q)}\equiv\int \prod_{i=1}^N d\tq_i^* d\tq_i d\qbar_i dq_i\;
\Exp\left[-\qbar Xq-\qbar A\tq-\tq^\dagger B q-\tq^\dagger Y\tq\right]\,,
\end{eqnarray}
where $X$ corresponds to the sea-quark Dirac operator and $Y$ to the
ghost-quark Dirac operator.  Here the variables $q_i$, $\qbar_i$,
$i=1,\dots,N$ are independent Grassmann-valued (fermionic) variables,
and the variables $\tq_i$, are bosonic, $c$-number variables.  The
matrices $X$ and $Y$ contain $c$-number entries, while the matrices $A$
and $B$ are Grassmann-valued.  Of course, the matrices $A$ and $B$
vanish in mixed-action QCD.  But, in order to study the possible
occurrence of a mixed sea-ghost condensate, one chooses (appropriate
entries of) $A$ and $B$ nonzero, and then one takes the limit of $A$ and
$B$ to zero after the volume has been taken to infinity.

Let us first recall the standard case for which also the $\tq_i$ are
fermionic.\footnote{In that case, the $q^\dagger_i$, like the $\qbar_i$,
  are independent fermionic variables as well.}
An example is the
formation of an Aoki condensate for two flavors of Wilson fermions.  In
this case we would take $X=Y$ equal to the hermitian Wilson--Dirac
operator, and we may take $A_{ij}=m_A\d_{ij}$,
$B_{ij}=-m_B\d_{ij}$.\footnote{For $m_A=m_B$ this corresponds to an Aoki
  condensate pointing in the $\s_2$ direction in isospin space.}
Without loss of generality, we may assume that $X$ has been
diagonalized,
\begin{equation}
\label{matrices}
X_{ij}=x_i\d_{ij}\ .
\end{equation}
We then find that
\begin{equation}
\label{standard}
\sum_{i=1}^N\langle\qbar_i\tq_i\rangle^{(q)}\equiv
-\frac{\partial}{\partial m_A}
Z^{(q)} = -Z^{(q)}\sum_{i=1}^N\frac{m}{x_i^2+m^2}\,,
\end{equation}
where in the last step we took $m_A=m_B=m$, and where
$\langle\dots\rangle^{(q)}$ is the unnormalized expectation value with
partition function $Z^{(q)}$.  After averaging over the gauge field and taking
the infinite-volume limit, followed by the limit $m\to 0$,
one finds a nonvanishing condensate if and only if there is a nonzero density
of near-zero modes \cite{Banks:1979yr}, because
\begin{equation}
\label{deltafunction}
\lim_{m\to 0}\frac{m}{x^2+m^2}=\pi\d(x)\,.
\end{equation}

Now let us return to the case in which the $\tq_i$ are bosonic.
The entries of $A$ and $B$ are now fermionic, and we can, in fact,
use this to work out what happens in any basis.
Taking $A_{ij}=\a\d_{ii_0}\d_{jj_0}$ for some fixed values of $i_0$ and $j_0$,
the integral~(\ref{Z}) now evaluates to
\begin{eqnarray}
\label{ZGr}
Z^{(q)}&=&\Sdet\left( \begin{array}{cc} X&A \\ B&Y \end{array} \right)
=\det\left(X-AY^{-1}B\right)/\det(Y)\\
&=&\det(X) \exp\left(- \tr(X^{-1} A Y^{-1} B)\right)/\det(Y)\nn\\
&=&
\det(X)\left(1 - \tr(X^{-1} A Y^{-1} B)\right)/\det(Y)\ ,\nn
\end{eqnarray}
and we find (taking Grassmann derivatives to be left derivatives)
\begin{equation}
\label{Grassmann}
\frac{\partial Z^{(q)}}{\partial\a}=-\left(Y^{-1} B X^{-1}\right)_{j_0i_0}
\det(X)/\det(Y)\ .
\end{equation}
It is clear  that in this case, in contrast to the standard case
reviewed above, no nonvanishing value can occur for
a Grassmann-valued condensate in the limit that $B\to 0$.
The basic reason is that we can always
expand the condensate in terms of the components of $A$ and $B$,
resulting in a finite polynomial in those components.
When we take $A$ and $B$ to zero, the corresponding
Grassmann-valued condensates will thus always vanish.
We conclude that when we consider
the vacuum expectation value for the field $\S$ in Eq.~(\ref{Sigma}),
we can set $\omega=\omegabar=0$.

Two more comments are appropriate.
First, one might wonder what would happen if
one takes the matrices $A$ and $B$ to be bosonic.
In this case, $Q=\qbar A\tq+\tq^\dagger Bq$
is Grassmann-valued, and it is easy to see, by expanding the exponent
in Eq.~(\ref{Z}) in terms of
$Q$, that in this case the partition function does not depend on $A$ and $B$
at all.  Condensates would thus trivially vanish.

Our second comment is that the same question can also be studied in ChPT.
In other words, one can assume that {\it a priori} $\omega$ and $\omegabar$
in Eq.~(\ref{Sigma}) do not vanish.  One then
finds that the equations of motion (for constant fields)
in the effective theory dictate that $\omega$ and $\omegabar$ vanish.
This is, of course, consistent with the QCD-based argument given above.

%============
%\begin{appendix}
%============
%=========================================
\section{\label{DeriveAnsatz} Proof of Eq.~(\ref{Sqform})}
%=========================================
Consider the matrix ({\it cf.} Eq.~(\ref{DefSq}))
\begin{equation}
\label{M}
\Sq = \left( \begin{array}{cc} \Sss & i\Ssv \\ i\Svs & \Svv
      \end{array} \right),\,
\end{equation}
where all the entries on the right-hand side are two-by-two matrices
with complex entries. As we have seen, $\Sq$ is an element of SU(4).
The unitarity constraint $\Sq \Sq^{\dagger} =\I$ provides constraints
on the submatrices:
\begin{subequations}
\begin{eqnarray}
\Sss\Sss^{\dagger} + \Ssv\Ssv^{\dagger} & = &\I\ ,\label{constra}\\
\Svs\Svs^{\dagger} + \Svv\Svv^{\dagger} & = &\I\ ,\label{constrb}\\
\Ssv\Svv^{\dagger}  &= & \Sss\Svs^{\dagger} \ .\label{constrc}
\end{eqnarray}
\end{subequations}
The effective theory is invariant under independent flavor rotations
in the sea and the valence sector. With $V_s\in $ U(2)$_{\rm sea}$
and $V_v\in $ U(2)$_{\rm val}$, $\Sq$ transforms into
\begin{equation}
\label{M'}
\Sq'=\left( \begin{array}{cc}
  V_s\Sss V_s^{\dagger} & iV_s\Ssv V_v^{\dagger} \\
  iV_v\Svs V_s^{\dagger} & V_v\Svv V_v^{\dagger}
\end{array} \right)\ .
\end{equation}
We now want to prove that this matrix can be brought into
the form~(\ref{Sqform})
if $\Svv=\r_v\I$, with $\r_v$ an arbitrary complex number.

First, from
Eq.~(\ref{constrb}) we conclude that
\begin{equation}
\label{C}
\Svs  =  \sqrt{1-|\rho_v|^2}\, C\ ,
\end{equation}
where $C \in {\rm U(2)}$.
We also find the bound $|\rho_v|\leq1$.
Equation~(\ref{constrc}) implies that
$\Ssv\Svv^{\dagger}\Svv\Ssv^{\dagger}
= \Sss\Svs^{\dagger}\Svs\Sss^{\dagger} $, hence
\begin{equation}
\label{BBd}
|\rho_v|^2\Ssv\Ssv^{\dagger}  =  (1 -|\rho_v|^2)\Sss\Sss^{\dagger}\ .
\end{equation}
Substituting this into Eq.~(\ref{constra}), we obtain
$\Sss\Sss^{\dagger}=\Svv\Svv^{\dagger}=|\rho_v|^2$,
and we may thus write $\Sss$ as
\begin{equation}
\label{A}
\Sss  =  \rho_v^*\,A\ ,
\end{equation}
where $A \in {\rm U(2)}$. From Eq.~(\ref{BBd}) we obtain
$\Ssv  = \sqrt{1-|\rho_v|^2}B$ , with, from Eq.~(\ref{constrc}),
$B=AC^\dagger$.  Performing a flavor transformation with $V_v=C^\dagger$,
we can thus write $\Sq$ as
\begin{equation}
\label{subS2}
\Sq=\left( \begin{array}{cc}
  \rho_v^* A & i\rhovperp\,A \\ i\rhovperp\I & \rho_v\I
\end{array} \right)\ ,
\end{equation}
where we introduced $\rhovperp = \sqrt{1-|\rho_v|^2}$.

The matrix $A$ can be written as $\exp(i\alpha_{\mu}\t^{\mu})$, where
$\t^\mu=(\I,{\vec\t})$. Hence we can define
$U=\exp(-i\alpha_{\mu}\t^{\mu}/2)$, with $AU^2=1$.
Performing a flavor rotation with $V_s= U$
and writing $U^{\dagger}=e^{i\phi}V$ with $V\in$~SU(2) simplifies $\Sq$ to
\begin{equation}
\label{subS5}
\Sq=e^{i\phi} \left( \begin{array}{cc}
  \rho_v^* e^{i\phi}V^2 & i\rhovperp V \\
  i\rhovperp V & \rho_ve^{-i\phi}
\end{array} \right)\ .
\end{equation}
Finally, the SU(2) matrix can be diagonalized by a flavor transformation
$V_s=V_v$ such that
\begin{equation}
\label{diagV}
V\to  D \,=\, \exp(i\theta\t_3/2)\ .
\end{equation}
Absorbing the phase $e^{-i\phi}$ into $\rho_v$, we arrive at
 \begin{equation}
\label{subS7}
\Sq=e^{i\phi} \left( \begin{array}{cc}
  \r_v^*D^2 & i\rhovperp D \\ i\rhovperp D & \r_v
\end{array} \right)\ .
\end{equation}
This is precisely Eq.~(\ref{Sqform}).   Because $\Sq\in\mbox{SU(4)}$,
it follows that $\det(\Sq)=e^{4i\phi}=1$, and thus that $\phi = 0$~mod~$\pi/2$.

\end{document}